\documentclass[twocolumn,graphics,prb,aps,floats,showpacs]{revtex4}
\usepackage{graphicx}
\usepackage{amsmath}
\usepackage{multirow}

\begin{document}

\title{RF bifurcation of a Josephson junction : microwave embedding circuit
requirements }
\author{V.E. Manucharyan}
\author{E. Boaknin}
\author{M. Metcalfe}
\author{R. Vijay}
\author{I. Siddiqi}
\thanks{Present address: Department of Physics, University of California,
Berkeley, California 94720}
\author{M. Devoret}
\affiliation{Department of Applied Physics, Yale University, New
Haven, CT 06511, USA}
\date{\today }

\begin{abstract}
A Josephson tunnel junction which is RF-driven near a dynamical bifurcation
point can amplify quantum signals. The bifurcation point will exist robustly
only if the electrodynamic environment of the junction meets certain
criteria. In this article we develop a general formalism for dealing with
the non-linear dynamics of Josephson junction embedded in an arbitrary
microwave circuit. We find sufficient conditions for the existence of the
bifurcation regime: a) the embedding impedance of the junction need to
present a resonance at a particular frequency $\omega _{R}$, with the
quality factor $Q$ of the resonance and the participation ratio $p$ of the
junction satisfying $Qp\gg 1$, b) the drive frequency should be low
frequency detuned away from $\omega _{R}$ by more than $\sqrt{3}\omega
_{R}/(2Q)$.
\end{abstract}

\pacs{85.25.Cp, 84.30.Le, 84.40.Az, 84.40.Dc}
\maketitle

\bigskip \bigskip \bigskip \bigskip






\section{Introduction}

Amplifying very small electrical signals is an ubiquitous task in
experimental physics. In particular, cryogenic amplifiers working in the
microwave domain found recently a growing number of applications in
mesoscopic physics, astrophysics and particle detector physics\cite%
{HEMT,Squid_amp_darkmatter}. We have recently proposed\emph{\ }\cite{JBA,
Dynamical_switching_of_JJ}\emph{\ } to use the dynamical bifurcation of a
RF-biased Josephson junction as a basis for the amplification of quantum
signals. A bifurcation phenomenon offers the advantage of displaying a
diverging susceptibility which can be exploited to maximize the amplifier
gain without necessarily sacrifying its bandwidth. Among all very low noise
and fast solid-state microwave devices, the Josephson junction distinguishes
itself by offering strongest non-linearity combined with weakest
dissipation. However, these characteristics are not by themselves
sufficient. The electrodynamic environment of the junction must also satisfy
a certain number of conditions in order for a controllable and minimally
noisy operation to be possible. In the recent Josephson bifurcation
amplifier experiments\cite{JBA, Dynamical_switching_of_JJ}, the junction was
shunted by a lumped element capacitor. A large capacitance had to be
fabricated very close to the junction to minimize parasitic circuit
elements, at the cost of severe complexity of patterning and thin-film
deposition. It would be very beneficial experimentally to simply embed the
Josephson junction in a planar superconducting microwave resonator. The aim
of this article is to establish theoretically the requirements that need to
be imposed on the embedding impedance of the junction in order to obtain a
bifurcation whose characteristics are suitable for amplification.

The article is organized as follows: after having briefly indicated the
connection between a bifurcating dynamical system and amplication, we review
the simplest non-linear dynamical system exhibiting the type of bifurcation
we exploit, namely the Duffing oscillator. We then describe the parameter
space of the oscillator, focussing on the neighborhood of the first
bifurcation and discussing why this is the most useful region. Having laid
the general framework for the analysis of our problem, we then consider the
simplest practical electrical implementation of the Duffing oscillator, a
Josephson junction biased by an RF source through an arbitrary microwave
circuit. The notion of embedding impedance is introduced. For concreteness,
we first examine the particular cases where the embedding impedance
corresponds to simple series or parallel LCR circuits. This allows us to
formulate the conditions under which the resulting non-linear electrical
system can be mapped into the Duffing model. We then examine the arbitrary
impedance case, finding that it must correspond to that of a resonator with
an adequate quality factor. We end the article by discussing possible
detailed experimental implementations of resonators and a concluding summary.

\section{Amplifying with the bifurcation of a driven dynamical system}

Amplification using a laser, a maser or a transistor is based on energizing
many microscopic systems, like atoms in a cavity or conduction electrons in
a channel, each one being weakly coupled to the input signal. The overall
power gain of the system, which is determined by the product of the number
of active microscopic systems and their individual response to the input
parameter, can be quite substantial. However, noise can result from the lack
of control of each individual microscopic system. This article explores
another strategy for amplification which involves a single system with only
one very well controlled collective degree of freedom, which is driven to a
high level of excitation. Here, the input signal is coupled parametrically
to this system and influences its dynamics. The best known device exploiting
this strategy is the SQUID\cite{SQUID} but other devices of the same type
have been proposed \cite{SSET}. Let us discuss the general question of the
gain (ratio of output to input) in such a system.

A driven dynamical system such as a SQUID is governed by a force equation
which, quite generally, can be written as

\begin{equation}
\mathcal{F}\left( \ddot{X},\dot{X},X,a\right) =F_{ext}\left( t\right)
\label{general_amplification}
\end{equation}

where $X$ is the system coordinate, $F_{ext}\left( t\right) $ is a periodic
external drive pumping energy in the system, $a$ a parameter of the system
and $\mathcal{F}$ a function describing its dynamics which is necessarily
dissipative since information is flowing away to the next stage of
amplification. We are interested in steady state solution of Eq. (\ref%
{general_amplification}), in which the energy flowing in from the source is
balanced by the energy losses. In the example of the rf SQUID, $X$ is the
total flux through the SQUID loop, $a$ the signal flux, and $F$ the external
driving flux with a frequency in the MHz range. For the dc SQUID\cite%
{RF_SQUID}, the frequency of the external drive current is $0$ and $X$ is
the common mode phase difference while $a$ is the flux through the loop
formed by the junctions. In this article we also consider a
Josephson-junction-based device like a SQUID, but it is driven by a rf
signal at microwave frequencies to increase speed and does not have
intrinsic dissipation.

When we use the dynamical system as an amplifier, we are linking the input
and output signals to the parameters $a$ and variable $X$, respectively.
Specifically, the signal $s(t)$ at the input of the amplifier induces a
variation $\delta a=\lambda s$. For a small input, the output $S\left(
t\right) $ of the amplifier will depend linearly on the modification of $X$:
$S\left( t\right) =\mathcal{L}\{X(t)-X(t)_{s=0}\}$. Since we are looking for
a maximal signal gain $S/s$, it is natural to find an operating point where
a small change in the parameter $a$ is going to induce a large change in the
dynamics of the system, provided we can keep all other parameters constant.
The largest susceptibility is found at a saddle-node bifurcation point and
it is in the neighborhood of such points that we will operate the amplifier.
The saddle-node bifurcation occurs when the drive parameters exceed certain
critical values. Previously, Yurke \textit{et al}. \cite{Yurke_review} have
studied Josephson systems mostly in the regime beneath these critical
values. Here, we consider similar systems, but we exploit instead the
bistable regime \emph{beyond} the critical values and the large
susceptibilities accompanying it. In the next section, we examine a simple
model exhibiting such a saddle-node bifurcation phenomenon.

\section{One minimal model for a bifurcating non-linear system: the Duffing
oscillator}

One of the most minimal model displaying the bifurcation phenomenon needed
for amplification is a damped, driven mechanical oscillator with a restoring
force displaying cubic non-linearity. The equation of this model, often
called the Duffing linear+cubic oscillator~\cite{Bogolyubov_Mitropolskii,
Nayfeh, Landau_Lifshitz}, is

\begin{equation}
m\ddot{X}+\gamma \dot{X}+kX(1-\nu X^{2})=F\cos \omega _d t+F_{N}\left(
t\right)  \label{Duffing_mechanical}
\end{equation}

\begin{figure*}[tbp]
\resizebox{\linewidth}{!}{%
\includegraphics{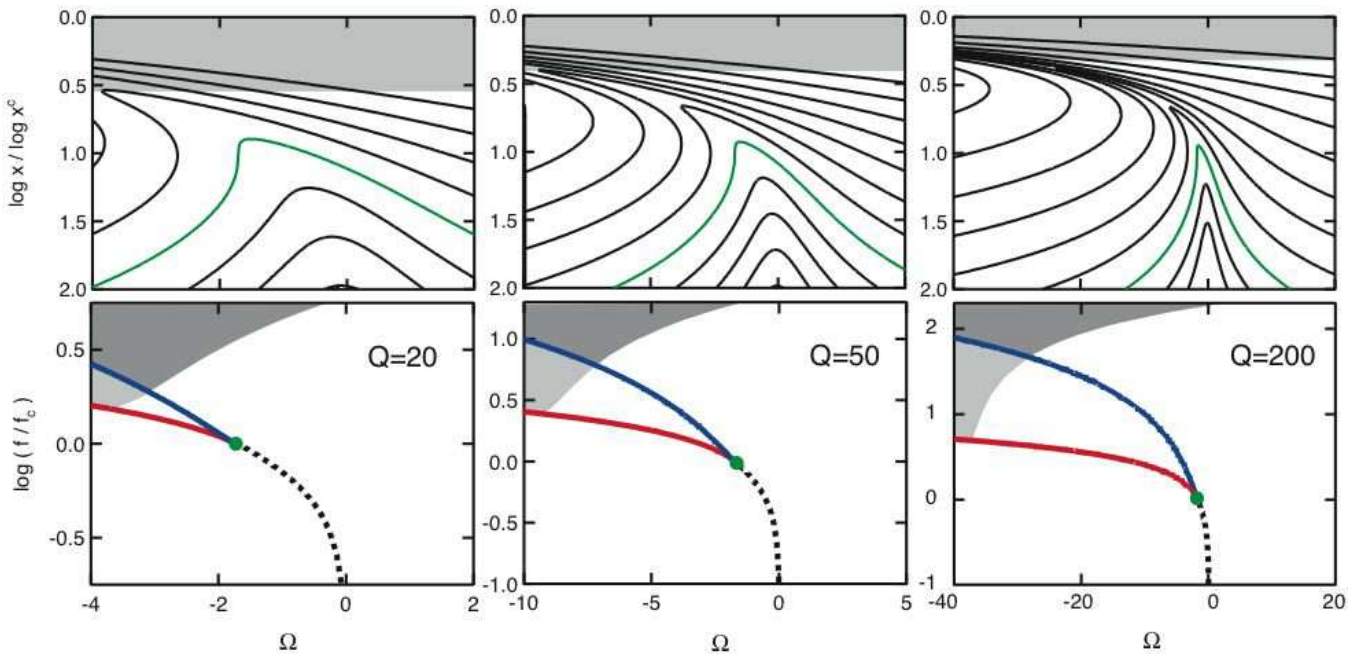}}
\caption{\textit{Upper panels:} Response of the Duffing linear+cubic
oscillator (see Eq.~\protect\ref{Duffing_mechanical}) as a function
of the dimensionless detuning $\Omega $, for different values of the
dimensionless driving amplitude $f$. The three panels correspond to
three different values of the small oscillation quality factor
$Q=20$, $50$, and $200$ and is shown over the same reduced frequency
range $0.9<w<1.05$. The response for the critical drive amplitude
$f_{c}$, where the curve presents a single point with diverging
susceptibilities $\partial {\bf x}/\partial \Omega $ and $\partial
{\bf x} /\partial f$, is shown in green. From bottom to top, the
increments in the drive amplitude $f$ are 2~dB (left), 2~dB (center)
and 3~dB (right). In the grey region, the oscillation amplitude is
taking place in the strongly non-linear regime (see text) and the
curves, which are calculated within the weak non-linear hypothesis,
are not to be trusted. \textit{Lower panels:} Stability diagram for
the dynamical states corresponding to the system response shown in
the upper panels. The y-axis is the drive amplitude $f$, scaled by
the critical amplitude $f_{c}$. The blue ($f_{B}(\Omega )$) and red
($f_{\bar{B}}(\Omega )$) lines delimit the region of bistability for
the system and correspond to the points of diverging susceptibility
which are visible in the upper panels. The two curves meet at the
critical point shown
by a green dot, whose position is determined by the critical detuning $%
\Omega _{c}=-\protect\sqrt{3}$ and drive amplitude $f_{c}$. The dashed line (%
$f_{ms}(\Omega )$) "continues" the blue and red lines and corresponds to the
points of maximal susceptibility with respect to driving force. The grey
regions correspond to that in the upper panels. Note that the "trusworthy"
domain of bistability increases monotonously with the quality factor $Q$.}
\label{fig1}
\end{figure*}

\bigskip

\bigskip

where $X$ is the position coordinate of the mechanical degree of freedom, $m$
its mass, $\gamma $ its damping constant, $k$ the stiffness constant of the
restoring force, $\nu $ the non-linearity parameter. The right handside
parameters $F$ and $\omega $ are the amplitude and angular frequency of the
driving force, respectively. For completeness, we have also added on the
right handside a noise force term $F_{N}\left( t\right) $ whose presence is
imposed by the fluctuation-dissipation theorem. It defines, through its
correlation function, a thermal energy scale for the problem, but in the
following sections, we are going to assume that this scale can be made much
smaller than all the other scales in the problem. The effect of fluctuations
will be treated in a latter article.

This simplification made, we can rescale the problem, changing the position
and time coordinates, obtaining in the end a three-parameter model

\begin{equation}
\ddot{x}+\frac{\dot{x}}{Q}+x(1-x^{2})=f\cos w\tau
\label{dimensionlessduffing}
\end{equation}

where the dot now refers to differentiation with respect to the rescaled
time $\tau $. The following equations express the rescaled quantities in
terms of the original ones: $\omega _{0}=\frac{1}{\sqrt{m/k}}$, $\tau
=\omega _{0}t$, $w=\frac{\omega _{d}}{\omega _{0}}$, $Q=\frac{\sqrt{mk}}{%
\gamma }$, $x=\sqrt{\nu }X$, $\ f=\frac{\sqrt{\nu }}{m}F$.

For reasons which will become clear later, we want to consider the low
damping limit of such a model and we suppose that the quality factor
satisfies $Q\gg 1$.

In the weak non-linear regime, i.e. $x\left( \tau \right) \ll 1$, the
frequency $w_{0}$ of natural oscillations ($f=0$) decreases with the
oscillation energy $u=\frac{\dot{x}^{2}}{2}+\frac{x^{2}}{2}-\frac{x^{4}}{4}$
as

\begin{equation}
w_{0}=1-\frac{3u}{2}+O\left( u^{2}\right)  \label{freqvsener}
\end{equation}

This corresponds to the softening of the restoring force as the amplitude of
the oscillatory motion increases.

The bifurcation phenomenon can be described crudely as follows: if we drive
the system below the small oscillation resonance frequency, increasing the
drive amplitude slowly, the resulting oscillation will be very small at
first. However, at a certain drive strength, the system becomes unstable and
tends to switch to a high amplitude oscillation where it can better meet the
resonance condition.

In a quantitative treatment of the weak non-linear regime, we seek a
solution involving only the first harmonic of the drive frequency\cite%
{Bogolyubov_Mitropolskii}

\begin{equation}
x\left( \tau \right) =\frac{1}{2}\tilde{x}\left( \tau \right) \mathrm{e}^{i
w\tau }+c.c.  \label{expansion}
\end{equation}

where the time dependence of the complex harmonic amplitude $\tilde{x}\left(
\tau \right) $ is slow on the time scale $w^{-1}$. Retaining in the equation
only the terms evolving in time like $\mathrm{e}^{iw\tau }$and ignoring $%
\mathrm{d}^{2}\tilde{x}\left( \tau \right) /\mathrm{d}\tau ^{2}$, one finds
the following relation for $\tilde{x}(\tau )$\cite{Dykman_Krivoglaz80}:

\begin{equation}
2i\dot{\tilde{x}}=\left( -\frac{\Omega +i}{Q}+\frac{3}{4}|\tilde{x}%
|^{2}\right) \tilde{x}+f  \label{Duffing_imaginary}
\end{equation}

in which we have introduced the reduced detuning $\Omega $
\begin{equation*}
\Omega =2Q(w-1)
\end{equation*}%
The static solutions ($\dot{\tilde{x}}=0$) for the modulus ${\bf
x}=|\tilde{x}|$ of the fundamental amplitude can be obtained as a
function of the parameters $(\Omega ,f)$ for a given $Q$ by solving
the equation

\begin{equation*}
f^{2}=\left( \frac{\Omega ^{2}+1}{Q^{2}}-\frac{3\Omega }{2Q}{\bf x}^{2}+%
\frac{9}{16}{\bf x}^{4}\right) {\bf x}^{2}
\end{equation*}

The susceptibility is given by the implicit expression:

\bigskip

\begin{equation*}
\frac{\partial {\bf x}}{\partial f}=\left( \frac{\partial f}{\partial %
{\bf x}}\right) ^{-1}=\frac{\sqrt{\frac{(1+\Omega ^{2})}{Q^{2}}-\frac{%
3\Omega }{2Q}{\bf x}^{2}+\frac{9}{16}{\bf x}^{4}}}{\frac{(1+\Omega
^{2})}{Q^{2}}-\frac{3\Omega }{Q}{\bf x}^{2}+\frac{27}{16}{\bf
x}^{4}}
\end{equation*}%
In the upper panels of Fig.~\ref{fig1}, we show ${\bf x}$ as a
function of $\Omega $ for increasing values of $f$ and for
$Q=20,50,200$. For small drive, the curve is the familiar Lorentzian
response of an harmonic oscillator, displaying a maximum response on
resonance at $\Omega =0$ and a half-width at half-maximum (HWHM)
point at $\Omega =1$. As the drive strength is increased, the
resonance curve bends towards lower frequencies, an indirect
manifestation of Eq. (\ref{freqvsener}). There is a critical drive
$f_{c}$ at which appears for the first time a critical reduced
detuning $\Omega _{c}$
such that the susceptibility $\partial \mathbf{x}/\partial f$ diverges\cite%
{Landau_Lifshitz}. We call $\mathbf{x}_{c}$ the oscillation
amplitude at this critical point. Analytic
calculations\cite{Landau_Lifshitz, Dykman_Krivoglaz80} lead to

\begin{eqnarray}
f_{c} &=&\frac{2^{5/2}}{3^{5/4}}\frac{1}{\sqrt{Q^{3}}}  \notag \\
\Omega _{c} &=&-\sqrt{3} \\
\mathbf{x}_{c} &=&\frac{2^{3/2}}{3^{3/4}}\frac{1}{\sqrt{Q}}  \notag
\label{Duffing_criticalpoint}
\end{eqnarray}

To be consistent with our weak non-linear regime hypothesis, we must have $%
\mathbf{x}_{c}\ll 1$ which implies in turn $Q\gg 1$.

For drives $f>f_{c}$, the response curve $\mathbf{x}\left( \Omega \right) $
develops an overhanging part in which there are three possible values for $%
\mathbf{x}$ at each value of $\Omega $. The smallest and highest values
correspond to two metastable states with different oscillation amplitudes,
whereas the intermediate value correspond to an unstable state for the
system. We denote as $f_{B}\left( \Omega \right) $ and $f_{\bar{B}}\left(
\Omega \right) $ the boundaries of this bistability interval: $f_{B}$ is the
force at which the system, submitted to an increasing drive with a fixed
frequency will switch from the low to the high amplitude state. Starting
from this state and decreasing the amplitude of the oscillatory force, the
system will switch back to the low amplitude state at $f_{\bar{B}}$. This
possibility of the Duffing system to "bifurcate" between two different
dynamical states at $f_{B}\left( \Omega \right) $ and $f_{\bar{B}}\left(
\Omega \right) $ is the phenomenon we are exploiting for amplification and
whose electrical implementation is the main topic of this paper. It is easy
to see that any input parameter coupled to $k$ or $m$ in Eq. (\ref%
{Duffing_mechanical}) will induce variations of the line $f_{B}\left( \Omega
\right) $. Fixing the drive parameters in the vicinity of this line, very
small changes in the input will induce large variations in the oscillation
amplitude. The variations can be reversible if we chose a point to the right
of the critical point (continuous amplifier operation) or the variations can
be hysteretic if we chose a point to the left of the critical point (latched
threshold detector operation).

In the limit $Q\gg 1$, analytic calculations can be carried further and lead
to\cite{Dykman_Krivoglaz80}

\begin{equation}
\frac{f_{B,\bar{B}}\left( \Omega \right)
}{f_{c}}=\frac{1}{2}\frac{\Omega ^{3/2}}{\Omega _{c}^{3/2}}\left[
1+3\frac{\Omega _{c}^{2}}{\Omega ^{2}}\pm \left( 1-\frac{\Omega
_{c}^{2}}{\Omega ^{2}}\right) ^{3/2}\right] ^{1/2}
\end{equation}

We define $f_{ms}(\Omega )$ as the line of maximum susceptibility $\partial
\mathbf{x}/\partial f$ on the low-frequency side of the resonance curve. It
defines the line of highest amplification gain below the bifurcation regime
\cite{susceptibility_footnote}. Its expression is given by:

\begin{equation}
\frac{f_{ms}(\Omega )}{f_{c}}=\frac{3^{1/2}}{2}\frac{\Omega ^{1/2}}{\Omega
_{c}^{1/2}}\left[ \frac{1}{3}\left( \frac{\Omega }{\Omega _{c}}\right) ^{2}+1%
\right] ^{1/2}
\end{equation}

The susceptibility on the high frequency side of the critical point diverges
as:

\begin{equation*}
\left. \frac{\partial \mathbf{x}}{\partial f}\right\vert _{\Omega \,\sim
\,\Omega c}=\frac{Q}{\Delta \Omega }
\end{equation*}%
where $\Delta \Omega $ is defined by $\Omega =\Omega _{c}-\Delta \Omega $
and $\Omega _{c}\gg \Delta \Omega $.

In the lower panels of Fig.~\ref{fig1}, we plot the bifurcation forces $f_{B}
$ (blue) and $f_{\bar{B}}$ (red), and $f_{ms}$ (dashed) normalized to the
critical force as a function of the reduced drive frequency $\Omega $. Note
that the lines representing $f_{B}(\Omega )$, $f_{\bar{B}}(\Omega )$ and $%
f_{ms}$ in the parameter space $(\Omega ,f/f_{c})$ are independent of the
parameters of Eq. (\ref{Duffing_mechanical}) and can be deemed "universal".

The dynamical critical point $(\Omega =\Omega _{c},f/f_{c}=1)$ is found at
the junction between the dashed line and the two bifurcation lines. One can
develop an analogy between the parameter space $(\Omega ,f/f_{c})$ and the
phase diagram of a fluid undergoing a liquid-vapor transition, the dynamical
critical point corresponding to the critical point beyond which vapor and
liquid cannot be distinguished by a transition (supercritical fluid regime),
and the bifurcation lines corresponding to the limit of stability of the
supercooled vapor and superheated fluid on either side of the 1st order
transition line (spinodal decomposition phenomenon).

\subsection{Weak and strong nonlinear regimes for the simple Duffing equation%
}

Let us now further discuss the small amplitude condition $\mathbf{x}\ll 1$,
which is necessary for the above results to hold. In Appendix A we show that
as long as

\begin{equation}
\frac{3}{2w^{2}}\mathbf{x}^{2}<1;~1/2<w<1  \label{Weak_non_linear}
\end{equation}

the Duffing model has stationary solution of the form

\begin{equation*}
x\left( \tau \right) =\frac{1}{2}\sum_{k=1}^{\infty }\left[ \tilde{x}_{2k-1}%
\mathrm{e}^{i\left( 2k-1\right) w\tau }+c.c.\right]
\end{equation*}

with the coefficients $\tilde{x}_{2k-1}$decreasing with the order $k$. Only
odd multiple of the drive frequency thus appear in this series. The first
harmonic coefficient $\tilde{x}_{1}$ is given by the stable solution of Eq. (%
\ref{Duffing_imaginary}) in the limit $|\tilde{x}_{1}|\ll 1$. Inequalities (%
\ref{Weak_non_linear}) defines rigourously the weak non-linear regime.

By contrast, in the strong non-linear regime $\frac{3}{2w^{2}}\mathbf{x}%
^{2}>1$, even harmonics start to proliferate as the oscillation amplitude
increases, leading eventually to chaotic behavior~\cite{Duffing_stability1,
Duffing_stability2, Duffing_stability3}. It is important to note that the
SQUID does not avoid this regime, even if, in general, strong dissipation
prevents fully developed chaos in this device.

Wanting at all cost to minimize noise in our use of this dynamical system
for amplification, we want to avoid the strong non-linear regime. Keeping in
mind that we are going to work with a small detuning $\left( \omega -\omega
_{0}\right) /\omega _{0}$, a conservative boundary separating the weak from
the strong non-linear regime can be introduced in parameter space by
requiring

\begin{equation}
\mathbf{x}<0.5  \label{condition1}
\end{equation}

In the lower panels of Fig.~\ref{fig1}, the grey region corresponds
to condition (\ref{condition1}) being violated for at least one of
the oscillation states. A lighter shade of grey marks the hysteretic
region between $f_{B}(\Omega )$ and $f_{\bar{B}}(\Omega )$ to
indicate that the low amplitude state does not violate
(\ref{condition1}) while the high amplitude does.

Note that in the reduced parameter space $(\Omega ,f/f_{c})$, the line $%
f_{s}(\Omega )$ corresponding to condition (\ref{condition1}) has a rather
drastic dependence on $Q$, in contrast with the other lines. Of course, if
we would plot the stability boundary lines in the absolute parameter space $%
\left( \omega ,F\right) $, the line corresponding to (\ref{condition1})
would be fixed while the critical point $\left( \omega _{c},F_{c}\right) $
would strongly depend on $Q$.

Whatever the representation, the important message which arises from the
stability diagram is that the amount of "real estate" in parameter space
that can be used for bifurcation amplification increases with $Q$. All
points along the $f_{B}(\Omega )$ line located between $f_{c}$ and $%
f_{s}(\Omega )$ are potentially useful. By realizing a large enough $Q$, one
can always "buy" the necessary amount of "real estate" in the stability
diagram, irrespectively of the value of the other parameters of the system.
Of course, higher $Q$ will tend to lower the bandwidth, but we can
compensate this effect by increasing the operating frequency of the device.

\section{The RF-biased Josephson junction}

\begin{figure}[tbp]
\resizebox{60mm}{!}{\includegraphics{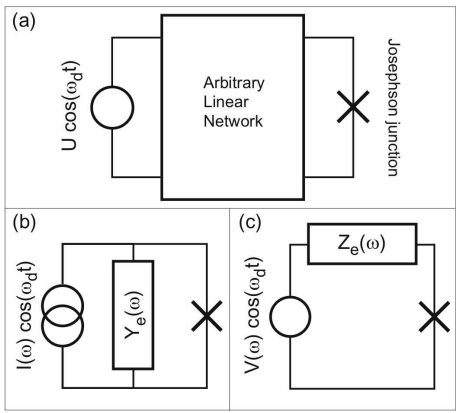}}
\caption{(a) Schematic of a Josephson junction RF-biased though an arbitrary
electric quadrupole containing only linear components. The Thevenin (b) and
Norton (c) representations of the circuit of (a) shows the dipole seen by
the junction. Note that the new source amplitude may now depend on
frequency. }
\label{arbcircuit}
\end{figure}
\noindent

We now apply these general considerations to the practical case of a
Josephson junction (JJ) biased by an RF source. At RF frequencies, the
impedance of the biasing circuit cannot be taken either as zero (ideal
voltage source) or infinite (ideal current source). In general, as depicted
by Fig.~\ref{arbcircuit}a, the biasing circuitry should be modeled as a
specific, frequency-dependent linear quadrupole connecting the Josephson
element of the junction to an ideal voltage source (the other circuit
element of the junction, its linear capacitance, has been lumped in this
quadrupole). Taking the point of view of the junction and applying either
the Norton or Thevenin representation, we arrive at the circuits of Fig.~\ref%
{arbcircuit}b and c respectively, where $Y_{e}(\omega )$ and $Z_{e}(\omega )$
are the admittance and impedance of the quadrupole seen from the junction.
In this transformation of the problem, the amplitude of the source may now
depend on frequency, but we ignore this complication for simplicity. It
turns out that in the cases of interest, either the Norton or Thevenin (but
not both) representions will satisfy this hypothesis adequately.

The Josephson electrical element, represented by a cross in our schematics,
is defined by its constitutive equation

\begin{equation}
I(t)=I_{0}\sin \left( \Phi _{J}\left( t\right) /\varphi _{0}\right)
\label{JJ_constitutive}
\end{equation}%
involving the generalized flux defined by $\Phi _{J}=\int_{-\infty
}^{t}V(t^{\prime })dt$\cite{Josephson_relations} and where $\varphi
_{0}=\hbar /2e$ is the reduced flux quantum. Here $V(t)$ and $I(t)$ are
respectively the voltage across the junction and current through it.

Comparing Eq. (\ref{JJ_constitutive}) with the constitutive equation of an
inductance

\begin{equation*}
I(t)=\frac{1}{L}\Phi \left( t\right)
\end{equation*}

we understand why the quantity $L_{J}=\left( \left. \frac{\partial I}{%
\partial \Phi _{J}}\right\vert _{\Phi _{J}=0}\right) ^{-1}=\varphi
_{0}/I_{0} $ is referred to as the effective Josephson inductance.

Note that Eq. (\ref{JJ_constitutive}) provides a non-linear link between $%
\Phi _{J}$ and $I$ and is, in some abstract sense, analogous to the relation
between force and position for a non-linear restoring force. However, the
external circuit described by $Y_{e}(\omega )$ or $Z_{e}(\omega )$ also
participate in the restoring force and we have to go through one further
step in order to establish a link between our electrical system and the
Duffing oscillator model.

\subsection{Separation of the linear and nonlinear contributions of the
Josephson element}

It is useful to split the Josephson element into its purely linear and
nonlinear components. The linear contribution of the Josephson element is
the impedance $iL_{J}\omega =Z_J(\omega )$, and can be incorporated in the
biasing impedance. The non-linear contribution, however, can be defined only
by first referring to either the Norton or the Thevenin representation.

We thus expand Eq. (\ref{JJ_constitutive}) in two different ways

\begin{equation}
I(t)={\frac{1}{L_{J}}\Phi \left( t\right) }-{\frac{1}{6L_{J}\varphi _{0}^{2}}%
{\Phi }_{J}\left( t\right) ^{3}+O\left[ \Phi ^{5}(t)\right] }
\label{spider_decomposition_Norton}
\end{equation}

and

\begin{equation}
\Phi _{J}\left( t\right) =L_{J}I(t)+\frac{L_{J}^{3}}{6\varphi _{0}^{2}}%
I^{3}(t)+O\left[ I^{5}(t)\right]  \label{spider_decomposition_Thevenin}
\end{equation}

\begin{figure}[tbp]
\resizebox{60mm}{!}{\includegraphics{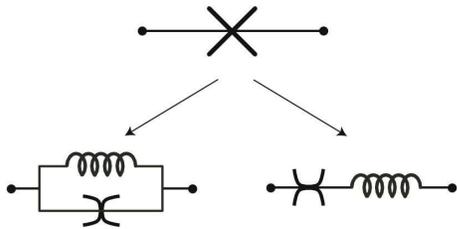}}
\caption{Two possible ways of separating the linear and nonlinear
contributions of a Josephson junction to a microwave circuit. Different
symbols were chosen for the nonlinear elements to suggest their different
current-voltage relations. The Josephson inductance in both cases has the
usual value $L_{J}=\protect\varphi _{0}/I_{0}$. }
\label{spiders}
\end{figure}

corresponding to these two representations, respectively. Relations~(\ref%
{spider_decomposition_Norton})~and~(\ref{spider_decomposition_Thevenin})
correspond respectively to a parallel and series combinations of a usual
linear inductance of value $L_{J}$ and a nonlinear element which is defined
by the higher order terms in the equations (see Fig~\ref{spiders}) and which
is also characterized by the parameter $L_{J}$. We will call these new
components \textit{parallel} nonlinear element (PNL) and \textit{series}
nonlinear element (SNL) respectively. They are represented by spider-like
symbols in Fig.~\ref{spiders}.

For the purpose of this paper, it will be sufficient to keep only the first
nonlinear term in each of the expansions above. The cut-line spider symbol
corresponds to the constitutive equation

\begin{equation}
I(t)=-\frac{1}{6L_{J}\varphi _{0}^{2}}\Phi _{J}\left( t\right) ^{3}
\label{para-spider}
\end{equation}%
while the spine-line spider symbol corresponds to the constitutive equation

\begin{equation}
\Phi \left( t\right) =\frac{L_{J}^{3}}{6\varphi _{0}^{2}}I^{3}(t)
\label{serial-spider}
\end{equation}

In the equation of our electrical system, these elements will lead to terms
analogous to the non-linear term in the Duffing Eq. (\ref{Duffing_mechanical}%
).

\begin{figure}[tbp]
\resizebox{60mm}{!}{\includegraphics{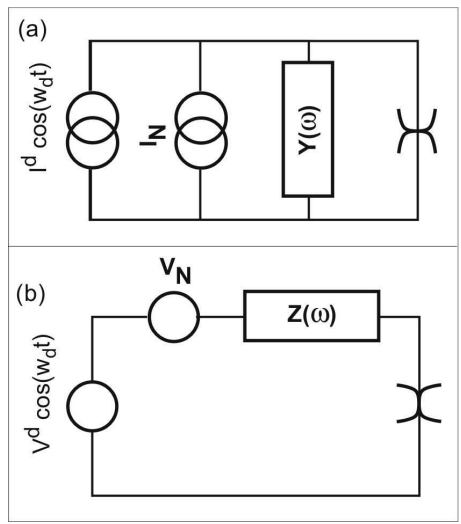}}

\caption{ Equivalent Norton (a) and Thevenin (b) representations of a linear
circuit driving a PNL and SNL Josephson element respectively. The
appropriate bias source is shown as either a parallel current source or a
series voltage source. }
\label{spiders_thevenin_norton}
\end{figure}

Fig.~\ref{spiders_thevenin_norton} shows the result of the complete
transformation of the initial circuit, in which the linear part of
the circuit is now described by the admittance $Y(\omega
)=Y_{e}(\omega )+Z_{J}^{-1}(\omega )$ in the Norton case, or by the
impedance $Z(\omega )=Z_{e}(\omega )+Z_{J}(\omega )$ in the Thevenin
case.

\noindent

\begin{figure}[tbp]
\resizebox{60mm}{!}{\includegraphics{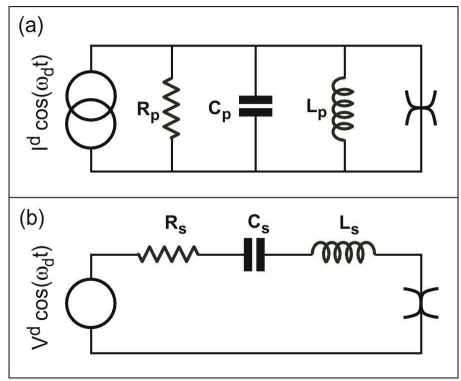}}
\caption{Josephson junction biased by (a) a parallel LCR circuit and
(b) a series LCR circuit. In the first case,
$L_{p}^{-1}=L_{pe}^{-1}+L_{J}^{-1}$ and in the second case,
$L_{s}=L_{se}+L_{J}$ where $L_{pe}$and $L_{se}$ are the inductances
contributed by the environment.} \label{LCRparaseries}
\end{figure}

Much qualitative insight can be gained from Fig.~\ref%
{spiders_thevenin_norton}a. Indeed, it is clear that, in order to have the
PNL participate and induce a significant nonlinear behavior, one needs the
current going through it to be large. For this to occur, the linear
admittance $Y(\omega )$ must be small. This will occur in the vicinity of a
resonance frequency of the linear part of the circuit, namely when $\mathrm{%
Im}[Y(\left\vert \omega \right\vert )]=0$. The same reasoning applies for
Thevenin representation (see Fig.~\ref{spiders_thevenin_norton}b) where the
voltage developed across the SNL must be large, i.e. in the vicinity of a
resonance where $\mathrm{Im}\left[ Z(\omega )\right] =0$ holds.

\subsection{The parallel and series LCR biasing circuits}

\begin{table}[t]
\begin{center}
\begin{tabular}{c|c|c|c|c|c}
\hline\hline
parameter & $\omega _{0}$ & $Q$ & $p$ & $f$ & $x$ \\ \hline\hline
mechanical & $\frac{1}{\sqrt{m/k}}$ & $\frac{\sqrt{mk}}{\gamma }$ & N.A. & $%
\frac{\sqrt{\nu }}{k}F$ & $\sqrt{\nu }X$ \\
parallel J+LCR & $\frac{1}{\sqrt{LC}}$ & $R\sqrt{\frac{C}{L}}$ & $\frac{L}{%
L_{J}}$ & $\sqrt{\frac{L}{6L_{J}}}\frac{LI^{d}}{\varphi _{0}}$ & $\sqrt{%
\frac{L}{6L_{J}}}\delta $ \\
series J+LCR & $\frac{1}{\sqrt{LC}}$ & $\frac{1}{R}\sqrt{\frac{L}{C}}$ & $%
\frac{L_{J}}{L}$ & $\sqrt{\frac{L_{J}}{6L}}\frac{V^{d}}{L\omega _{0}I_{0}}$
& $\sqrt{\frac{L_{J}}{6L}}\frac{\omega _{0}}{I_{0}}q$ \\ \hline\hline
\end{tabular}%
\end{center}
\caption{Table of correspondance between the parameters of the various
realizations of the generalized Duffing equation (see Eq.~(\protect\ref%
{Duffing_imaginary})) considered in this article, i.e. the mechanical
oscillator and the electrical oscillators based on a Josephson junction
biased either through a parallel or a series LCR circuit. The notion of
participation ratio $p$ has meaning only in the case of the Josephson
junction circuits in which the current through the junction cannot exceed
its critical current.}
\label{table1}
\end{table}

\begin{table*}[t]
\begin{center}
\begin{tabular}{l|l|l||l}
\hline\hline
& \multicolumn{2}{c||}{parameters at critical point} & \multicolumn{1}{c}{
response at critical point} \\ \hline
& \multicolumn{1}{c|}{frequency} & \multicolumn{1}{c||}{drive amplitude} &
\multicolumn{1}{c}{oscillation amplitude} \\ \hline\hline
mechanical & $\Omega _{c}\equiv 2Q\left( \frac{\omega _{c}}{\omega _{0}}%
-1\right) =-\sqrt{3}$ & $f_{c}=\frac{2^{5/2}}{3^{5/4}}\frac{1}{\sqrt{Q^{3}}}$
& ${{\bf x}_c=\frac{2^{3/2}}{3^{3/4}}\frac{1}{\sqrt{Q}}}$ \\
parallel J+LCR & $\Omega _{c}\equiv 2Q\left( \frac{\omega _{c}}{\omega _{0}}%
-1\right) =-\sqrt{3}$ & $I_{c}^{d}=\frac{8}{3^{3/4}}\left( \frac{1}{Q}\frac{%
L_{J}}{L}\right) ^{3/2}I_{0}$ & $I_{c}\equiv \delta _{c}I_{0}=\frac{4}{%
3^{1/4}}\sqrt{\frac{1}{Q}\frac{L_{J}}{L}}I_{0}$ \\
series J+LCR & $\Omega _{c}\equiv 2Q\left( \frac{\omega _{c}}{\omega _{0}}%
-1\right) =-\sqrt{3}$ & $V_{c}^{d}=\frac{8}{3^{3/4}}\left( \frac{1}{Q}\frac{L%
}{L_{J}}\right) ^{3/2}\omega _{c}\varphi _{0}$ & $I_{c}\equiv \omega
_{c}q_{c}=\frac{4}{3^{1/4}}\sqrt{\frac{1}{Q}\frac{L}{L_{J}}}I_{0}$ \\
parallel $Y(\omega )$ & $\frac{\mathrm{Im}{[Y(\omega _{c})]}}{\mathrm{Re}{%
[Y(\omega _{c})]}}=-\sqrt{3}$(implicit) & $I_{c}^{d}=\frac{8}{3^{3/4}}\left(
\frac{\mathrm{Re}{[}Y(\omega _{c})]}{|Y_{J}(\omega _{c})|}\right)
^{3/2}I_{0} $ & $I_{c}=\frac{4}{3^{1/4}}\sqrt{\frac{\mathrm{Re}[Y(\omega
_{c})]}{|Y_{J}(\omega _{c})|}}I_{0}$ \\
series $Z(\omega )$ & $\frac{\mathrm{Im}[Z(\omega _{c})]}{\mathrm{Re}{%
[Z(\omega _{c})]}}=-\sqrt{3}$(implicit) & $V_{c}^{d}=\frac{8}{3^{3/4}}\left(
\frac{{\mathrm{Re}[Z(\omega _{c})]}}{|Z_{J}(\omega _{c})|}\right)
^{3/2}\omega _{c}\varphi _{0}$ & $I_{c}=\frac{4}{3^{1/4}}\sqrt{\frac{\mathrm{%
Re}{[Z(\omega _{c})]}}{|Z_{J}(\omega _{c})|}}I_{0}$ \\ \hline\hline
\end{tabular}%
\end{center}
\caption{Values of system parameters (frequency, drive amplitude) and
response (amplitude of oscillation) at the critical point for the various
mechanical and electrical realizations of the Duffing system considered in
this article. In the two boxes marked "implicit", the critical frequency can
be obtained only by the solving the given equation.}
\label{table1b}
\end{table*}

We now consider the simplest cases where the environment of the non-linear
element -- described by either $Y\left( \omega \right) $ or $Z\left( \omega
\right) $ -- is either the parallel or the series LCR circuit (see Fig. \ref%
{LCRparaseries}). These circuits are defined by the inductances,
capacitances and resistances $L_{p}$, $C_{p}$, $R_{p}$ in the parallel case
and $L_{s}$, $C_{s}$, $R_{s}$ in the series case. From the discussion in the
last section, $L_{p}$ and $L_{s}$ include contribution of the Josephson
effective inductance: for the parallel case, $%
L_{p}^{-1}=L_{pe}^{-1}+L_{J}^{-1}$ and for the series case, $%
L_{s}=L_{se}+L_{J}$, where $L_{pe}$ and $L_{se}$ are the inductances of the
embedding circuit of the physical junction.

For the parallel case, the application of Kirchhoff's law to currents in all
the branches leads to the equation of motion

\begin{equation}
C_{p}\varphi _{0}\ddot{\delta}+\frac{\varphi _{0}}{R_{p}}\dot{\delta}+\frac{%
\varphi _{0}}{L_{p}}\delta \left( 1-\frac{L_{p}}{6L_J}\delta ^{2}\right)
=I^{d}\cos \omega t  \label{paraLCR}
\end{equation}

which is a strict analog of Eq.~(\ref{Duffing_mechanical}). Here we have
introduced the so-called gauge-invariant phase difference $\delta =\Phi
_{J}\left( t\right) /\varphi _{0}$. In Table \ref{table1} we show the
correspondance between the mechanical system and this parallel LCR system.
The associated critical parameters are given in Table \ref{table1b}. Note
that in this table the critical coordinates are referred to using the
current $I_{c}=\delta _{c}I_{0}$. To a good approximation, this is the
amplitude of the current through the junction at the critical point.

We now identify an important parameter which we call the parallel \textit{%
participation ratio}

\begin{equation}
p_{p}=L_{p}/L_J
\end{equation}

Together with the quality factor $Q$, it determines the ratio between the
current $I_{pc}$ at the dynamical critical point and the maximum Josephson
supercurrent $I_{0}$ (see Table \ref{table1b}). The participation ratio
measures the strength of the nonlinearity: a small participation ratio is
associated with a weak nonlinear term when $\delta \sim 1$ (see Eq.~(\ref%
{paraLCR})).

We now turn to the case of the series LCR circuit shown in Fig.~\ref%
{LCRparaseries}. Here, summing all the voltages across the elements of the
circuit, we arrive at another equation of motion given by

\begin{equation}
L\ddot{q}\left( 1+\frac{1}{2}\frac{L_J}{LI_{0}^{2}}~\dot{q}^{2}\right) +R%
\dot{q}+Cq=V^{d}\cos \omega t
\end{equation}

where $q$ is the total charge having flown through the junction up to time $%
t $. The cubic nonlinearity of this equation does not affect the restoring
force like in the Duffing equation but the mass of the particle.
Nevertheless, it is easy to show the same Duffing oscillator dynamics is
recovered at low drives within the single harmonic approximation (see Table~%
\ref{table1}). Like for the parallel case, the critical amplitude in Table %
\ref{table1b} is given in a terms of a current $I_{sc}=\omega _{c}q_{c}$,
which also here corresponds to the amplitude of the current through the
junction at the critical point.

Again, we define the series \textit{participation ratio}
\begin{equation}
p_{s}=L_J/L
\end{equation}

which plays the exact same role as $p_{p}$ for the series case.

That the parallel and series case can be mapped into one another in the weak
non-linear regime (see Table \ref{table1}) is not an accident. Quite
generally, one can show that any linear oscillator equation to which is
added a cubic nonlinearity in any combination of $x$ and its derivatives ($%
\dot{x}$, $\ddot{x}$, etc.) will lead to, in the weak non-linear regime, the
same dynamics as that of Eq.~(\ref{Duffing_imaginary}).

The striking conclusion of this section on a lumped element resonator is
that even when the junction has a weak participation ratio, its
non-linearity is not really "diluted". It will still display a bifurcation
which can be employed for amplification, provided that the control of the
amplitude of the oscillatory drive meets a corresponding increase in
precision.

\subsection{General biasing circuit involving a resonator}

At the microwave frequencies where we wish to work, it is difficult to
implement a pure LCR circuit without substantial parasitic elements. In
practice, it will be easier to implement a distributed element resonator
built with section of transmission lines\cite{Zmuid,Wallraff}. However, as
we are going to demonstrate, the conclusions of the last section are robust
provided that the quality factor of the resonator is chosen adequately,
which is easily achievable with on-chip superconducting thin film coplanar
waveguides. We therefore now consider the general case of an arbitrary
impedance $Z\left( \omega \right) $ and use the Thevenin representation (see
Fig.~\ref{spiders_thevenin_norton}b). The extension of our results to the
case of an arbitrary admittance in the Norton representation will be
straightforward, using the set of simultaneous transformation: $Z(\omega
)\rightarrow Y(\omega )$, $Z_{J}(\omega )\rightarrow Y_{J}(\omega )$, SNL$%
\rightarrow $PNL.

We start by writing the equations of motion using Kirchoff's laws and
looking for a solution of the form $I(t)=\frac{1}{2}\tilde{I}(t)e^{i\omega
t}+c.c.$. By retaining only the drive harmonic terms and staying in the
limit of small oscillation amplitude where we can take the nonlinearity into
account at the lowest order, one finds an equation of the form (\ref%
{Duffing_imaginary}):

\begin{equation}
\sum_{n=1}^{\infty }i^{n}Z^{(n)}\tilde{I}^{(n)}(t)=\left( Z+\frac{1}{8}%
Z_{J}\left\vert \frac{\tilde{I}(t)}{I_{0}}\right\vert ^{2}\right) \tilde{I}%
(t)-V_{d}  \label{Thevenin_RWA_dynamics}
\end{equation}

Here , $Z=Z(\omega =\omega _{d})$, $Z_{J}=i\omega _{d}L_{J}$, $Z^{(n)}=\frac{%
\mathrm{d}^{n}Z(\omega )}{\mathrm{d}\omega ^{n}}|_{\omega =\omega _{d}} $
are the derivatives of $Z(\omega )$ taken at the drive frequency and $\tilde{%
I}^{(n)}=\frac{\mathrm{d}^{n}I(t)}{\mathrm{d}t^{n}}$. One can check that
when $\left\vert \tilde{I}/I_{0}\right\vert \ll \sqrt{|Z/Z_{J}|}$, i.e. when
nonlinearity can be neglected, the equation describes the complete response
of the system. In particular, the steady state oscillation amplitude is
given by $\tilde{I}=V_{d}/Z(\omega )$. The zeros of $Z(\omega )$ are thus
the complex resonant frequencies corresponding to the normal modes of
excitations of the linear system (A well-built resonator has only a sparse
set of zeros regularly distributed along the real frequency axis). The time
derivative term on the left-hand side of Eq. (\ref{Thevenin_RWA_dynamics})
accounts for the transient dynamics.

The regime of interest for amplification is when $\left\vert \tilde{I}%
/I_{0}\right\vert \ $approaches $\sqrt{|Z/Z_{J}|}$ for frequencies in the
vicinity of a zero of $Z(\omega )$. We then find an equation bearing a close
resemblance with Eq. (\ref{Duffing_imaginary}) for the first harmonic of the
Duffing oscillator amplitude. In the following, we will neglect all higher
derivatives $Z^{(n>1)}$.

It is important to note that while Eq. (\ref{Duffing_imaginary})
yields the bifurcation amplitude and drive as simple functions of
the detuning frequency $\Omega $, equation
(\ref{Thevenin_RWA_dynamics}) only leads to the amplitude
$\frac{\tilde{I}(t)}{I_{0}}$ and drives $V_{d}$ at the bifurcation
points as a function of the complex quantity $Z$ and $Z'$, which
themselves have
complicated dependence on $\omega_{d}$. Although the expressions for generalized quantities such as $%
f_{B}(\Omega )$, $f_{\bar{B}}(\Omega )$ and $f_{ms}(\Omega )$ are tractable,
they provide little insight. We therefore refrain from presenting them in
this paper.

Nevertheless, we can still identify conditions that $Z$ and $Z'$
need to satisfy in order for the bifurcation real estate to exist.
These conditions are obtained from Table \ref{table1b}$.$

\begin{figure}[tbp]
\resizebox{80mm}{!}{\includegraphics{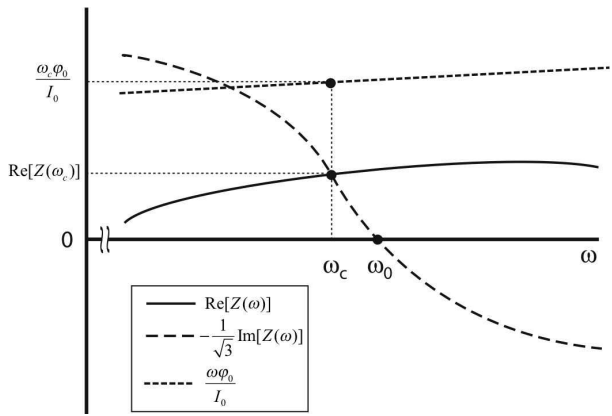}}
\caption{Sketch of the graphical determination of whether a circuit
characterized by impedance $Z(\protect\omega )$ (see text) will allow a
Josephson junction with critical current $I_{0}$ to bifurcate. The procedure
consists in i) plotting simultaneously $\mathrm{Re}[Z(\protect\omega )]$, $-%
\frac{1}{\protect\sqrt{{3}}}\mathrm{Im}[Z(\protect\omega )]$ and $Z_{J}(%
\protect\omega )=\protect\omega \protect\varphi _{0}/I_{0}$, ii) finding the
critical frequency $\protect\omega _{c}$ (if it exists) at the eventual
intersection $\mathrm{Re}[Z(\protect\omega )]=-\frac{1}{\protect\sqrt{3}}%
\mathrm{Im}[Z(\protect\omega )]$, and iii) verifying that the condition for
weak nonlinearity is satisfied at $\protect\omega _{c}$, namely, $Z_{J}(%
\protect\omega _{c})\gg \mathrm{Re}[Z(\protect\omega _{c})]$. }
\label{find_wc_procedure}
\end{figure}

Firstly, a critical drive frequency $\omega _{c}$ must exist such that the
impedance $Z(\omega _{c})$ satisfies the condition $\mathrm{Re}[Z(\omega
_{c})]=-\sqrt{3}$\textrm{Im}$[Z(\omega _{c})]$. A way to meet this condition
is for the embedding impedance to have at least one resonant frequency $%
\omega _{0}$ where \textrm{Im}$\left[ Z\left( \omega _{0}\right) \right] =0$%
. Since for a passive circuit\cite{Pozar}

\begin{eqnarray*}
\left. \frac{\mathrm{d}\text{\textrm{Im}}\left[ Z\left( \omega \right) \right] }{%
\mathrm{d}\omega }\right\vert _{\omega _{0}}
&=&\text{\textrm{Im}}\left[ Z^{\prime
}\left( \omega _{0}\right) \right] >0 \\
\mathrm{Re}[Z(\omega _{0})] &>&0
\end{eqnarray*}

a critical frequency can be found at

\begin{equation*}
\omega _{c}\simeq \omega _{0}-\frac{\mathrm{Re}[Z(\omega _{0})]}{\sqrt{3}%
\text{\textrm{Im}}\left[ Z^{\prime }\left( \omega _{0}\right) \right] +%
\mathrm{Re}\left[ Z^{\prime }\left( \omega _{0}\right) \right] }
\end{equation*}

assuming the impedance is a smooth function of frequency near the resonant
frequency. In Fig.~\ref{find_wc_procedure} we show how an exact solution for
the frequency $\omega _{c}$ can be obtained graphically from a measurement
of \textrm{Im}$\left[ Z\left( \omega \right) \right] $ and $\mathrm{Re}%
[Z(\omega )]$ in the neighborhood of $\omega _{0}$. The graphical
construction shows that the critical drive frequency needs to be located on
the low-frequency flank of a resonance of the bias circuit, as in the case
of the simple LCR circuit.

A second condition has to be fulfilled, however: the last line of Table \ref%
{table1b} indicates that, at the critical frequency, the real part \textrm{Re%
}$[Z(\omega _{c})]$ of the impedance must be much smaller than the effective
impedance $\omega _{c}\varphi _{0}/I_{0}$ = $L_{J}\omega _{c}$ of the
junction. The ratio of the RF\ critical current $I_{c}$ to the DC critical
current $I_{0}$ must be much smaller than one in order to fulfill the
condition of weak non-linearity. The smaller the $\mathrm{Re}[Z(\omega
_{c})]/|Z_{J}(\omega _{c})|$ ratio, the larger becomes the "real estate"
available for bistability in the stability diagram. This last consideration
sets the maximum amount of dissipation which can be tolerated in the
embedding impedance $Z_{e}\left( \omega \right) $.

It is useful to translate this last condition in a more practical language.
If we introduce the generalized quality factor $Q\left( \omega \right) $ of
the environmental impedance:

\begin{equation*}
Q\left( \omega \right) =\frac{\mathrm{Re}[Z_{env}(\omega )]}{\omega \text{%
\textrm{Im}}\left[ Z_{env}^{\prime }\left( \omega \right) \right] }
\end{equation*}%
we find that the condition on $I_{c}$ can be rewritten as

\begin{equation*}
\frac{8}{3^{1/4}}\sqrt{\frac{1}{Q\left( \omega _{c}\right) p}}<1
\end{equation*}

Thus, the higher the product of $Q$\textit{\ }and\textit{\ }$p,$ the greater
real estate for amplification.

\section{Further discussion}

The phenomenon of bistability has been recently observed in an electrical
circuit biasing a Josephson junction \cite%
{Dynamical_switching_of_JJ,janicelee}. Its operation as an amplifier has
been studied \cite{JBA} and used to measure a superconducting qubit \cite%
{JBA_and_qubit,lupascu}. The amplification near a bifurcation point is
limited by a stochastic dynamical escape process similar to the one observed
in DC biased JJ. This process is well studied theoretically in the classical
regime where the fluctuations stem from thermal noise \cite%
{Dykman_Krivoglaz79,Dykman_Krivoglaz80}. The quantum regime is not
as well understood but important results can be found in
\cite{Dykman_Smelyanski88, Dykman}. The circuit of Ref.
\cite{Dynamical_switching_of_JJ} is based on the simplest
realization of such an RF biased junction, namely, a parallel LCR
circuit. The resistance was provided by the $50~\Omega $ RF
environment, the capacitance was constructed lithographically on
chip, and the inductance came entirely from the junction itself.

Another approach is to use continuous circuit elements instead of discreet
ones such as L's, R's and C's. A resonator structure with low internal loss
can be used to bias the junction. Examples of these can range from a
waveguide resonator, a lithographic resonator (see for \textit{e.g.} \cite%
{Wallraff}), or any other resonator geometry. Any harmonic of such structure
can be used to bias the JJ. The key lies in finding the correct way to
couple the JJ. Let us discuss briefly a particular implementation, namely,
using a coplanar waveguide cavity resonator such as the ones recently used
to observe cQED in a solid state system \cite{Wallraff}. There, a junction
can be placed at the center of the transmission line, interrupting its
central conductor. The circuit analysis shows that this corresponds to the
case discussed above of a JJ biased by a series LCR circuit. We have
successfully fabricated and tested such devices and will discuss the results
in a later paper.

\section{Concluding summary}

In this paper we have presented and discussed the idea of harnessing, for
amplification purposes, the bifurcation phenomenon of a driven Josephson
junction embedded in a resonating circuit. We have first reviewed the basic
mathematics associated with the general saddle-node bifurcation phenomenon
using the simplest example of the mechanical Duffing oscillator. We have
then applied the corresponding formalism to a simple RF driven LCR
electrical oscillator incorporating a Josephson junction and found the
conditions for observing bistability and a diverging susceptibility. There
are two relevant parameters: the quality factor $Q$ and the participation
ratio. The latter quantity is necessarily smaller than unity and is a
measure of the strength of the nonlinearity provided by the JJ, relative to
the embedding circuit. We have found that the product of these two
parameters need to be much greater than one in order to provide a convenient
operating conditions for amplification. We have then shown that the same
condition applies to a more general resonating structure, with a simple
adaptation of the definition of the quality factor and participation ratio
to the vicinity of a resonance. Our prescription involves a relatively
precise analysis of the microwave circuit of the resonator but can easily be
followed with a varieties of geometries since microwave resonators can
easily be made to have large quality factors. Later generations of
amplifiers or detectors based on the bifurcation of superconducting
resonators should therefore benefit from the analysis presented in this
article.

\bigskip

\begin{acknowledgments}
\bigskip We would like to thank D. Esteve, D. Vion and R. Schoelkopf for
useful discussions. \bigskip
\end{acknowledgments}

\bigskip

\appendix

\section{Weak and strong non-linear regimes for a non-linear oscillator}

In this appendix we would like to examine the behavior of a driven nonlinear
oscillator beyond the simple framework developed in the main body of the
paper which considers that the circuit only responds at the frequency of the
driving force. We investigate the limits in which this hypothesis is
violated, specializing to the realistic case of a Josephson junction. Our
analysis relies heavily on the work developed in~\cite%
{Bogolyubov_Mitropolskii, Nayfeh, Duffing_stability1, Duffing_stability2,
Duffing_stability3}. We will consider the generalization of Eq. (\ref%
{Duffing_mechanical}) to an arbitrary nonlinear potential. To simplify the
notations and to avoid confusions with the main text, we use $X\rightarrow y$%
, $m\rightarrow 1$, replace $\omega _{d}\rightarrow \omega $ and drop the
noise term $F_{N}(t)$ from Eq.~(\ref{Duffing_mechanical}). The purely
non-linear force is written as $N(y)$ and we thus start with

\begin{equation}
\ddot{y}+\gamma \dot{y}+\omega _{0}^{2}y+N(y)=f\cos \omega t
\label{appendix_equation_of_motion}
\end{equation}

We will restrict our discussion to a nonlinear force $N(y)$ which is an
antisymmetric function of $y$ as this is the case relevant to the
nonlinearity provided by a Josephson junction. It will also simplify the
discussion.

A solution to Eq.~(\ref{appendix_equation_of_motion}) can be found in the
following form:

\begin{equation}  \label{appendix_generic_solution}
y(t) = y^{odd}(t, n) + \delta y (t, n)
\end{equation}

where function $y^{odd}(t,n)$ is a sum of a finite number $n$ of odd
harmonics

\begin{equation}
y^{odd}(t,n)=\frac{1}{2}\sum_{k=1}^{n}\tilde{y}_{2k-1}\mathrm{e}%
^{i(2k-1)\omega t}+c.c.  \label{appendix_x_tau_n}
\end{equation}

while $\delta y(t,n)$ is a correction to the expansion. Note that $n$ is
introduced here as an integer-valued parameter.

By assuming $\delta y(t,n)\ll y^{odd}(t,n)$ we will conduct a linear
analysis of this correction. Keeping $Q=\omega _{0}/2\gamma $ constant (see
Eq.~(\ref{appendix_equation_of_motion})), we find two different behaviors
for $\delta y(t,n)$ depending on the parameters $\{\omega ,f\}$ (their
values are discussed later below) :

\textbf{1.} $\delta y(t,n)$ is bounded in time and contains only \textit{odd}
harmonics of $\omega $ starting with $2n+1$. It follows that Eq.(\ref%
{appendix_x_tau_n}) is a good approximation for the stationary solution of
Eq.~(\ref{appendix_equation_of_motion}). The choice of $n$ is determined by
the required accuracy of the solution. The amplitudes of the stationary
oscillations $\tilde{y}_{k}$ are functions of drive frequency $\omega $ and
amplitude $f$ given by a system of nonlinear \textit{algebraic} equations in
terms of $N(y)$.

\textbf{2.} $\delta y(t,n)$ is unbounded in time. This would lead to the
breakdown of the validity of our linear analysis, or is at least an
indication that the form for the solution given in Eq.~(\ref%
{appendix_x_tau_n}) has to be modified. The instability of $\delta y(t,n)$
can be of two types:

\textbf{2a.} $\delta y(t,n)\sim \mathrm{e}^{\lambda t}\mathrm{e}^{i\omega
(2k-1)t}$, where $k\leq n$ is an integer and $\lambda >0$ is a Lyapunov
exponent. The instability of this type corresponds to the switching of the
oscillation amplitude from one stationary state to another at frequency $%
(2k-1)\omega $. This instability is of major interest to us as a resource
for amplification purposes. In particular, for $k=1$, a switching between
two stable states can occur.

\textbf{2b.} $\delta y(t,n)\sim \mathrm{e}^{\lambda t}\mathrm{e}^{i2k\omega
t}$, $k\leq n$ is an integer. This is a different instability phenomenon
because the solution contains growing \textit{even} harmonics which breaks
the symmetry of the nonlinearity $N(-y)=-N(y)$. It was shown to be a
precursor of chaotic behavior of a nonlinear oscillator, at least for the
Duffing case, where $N(y)\propto y^{3}$.

Let us fix $n$ and write down the system of nonlinear algebraic equations
that defines $\tilde{y}_k$, $k = 1, 2, ..., n$ by using the harmonic balance
method:

\begin{equation}  \label{appendix_N_stationary_amplitudes}
\begin{split}
\tilde{y}_{2k-1} \left[1 -(2k-1)^2\omega^2 + i (2k-1)\omega \gamma \right] +
\qquad \qquad \\
\tilde{N}_{2k-1}^n(\tilde{y}_1, ..., \tilde{y}_{2n-1}) = \frac{f}{2}
\delta_{1, 2k-1} \qquad
\end{split}%
\end{equation}

where $\delta_{1, 2k-1}$ is the Kronecker delta function. The complex
functions $\tilde{N}_{2m-1}^n(z_1, z_3, ..., z_{2n-1})$ are defined as:

\begin{equation}
\begin{split}
\tilde{N}_{2m-1}^{n}(z_{1},..,z_{2n-1})= \\
\int_{-\pi }^{\pi }N\left( \frac{1}{2}\sum_{k=1}^{n}z_{2k-1}\mathrm{e}%
^{i(2k-1)\theta }+c.c.\right) \mathrm{e}^{-i(2m-1)\theta }\,\frac{d\theta }{%
2\pi }
\end{split}
\label{appendix_definition_of_tilde(N)}
\end{equation}

The solution of Eq.~(\ref{appendix_N_stationary_amplitudes}) gives the
amplitudes $\tilde{y}_{2k-1}(f,\omega )$ as a function of drive amplitude
and frequency. Note that only the $k=1$ harmonic is driven directly by an
external force, while the higher harmonics feel the drive via the
nonlinearity.

The correction $\delta y(t,n)$ is obtained by subtracting the solutions to $%
y^{odd}(t,n)$ (see Eq.~(\ref{appendix_N_stationary_amplitudes})) from the
solution of the initial equation (Eq.~(\ref{appendix_equation_of_motion}))
and keeping the linear terms in $\delta y$. This leads to:

\begin{equation}  \label{appendix_general_correction}
\begin{split}
\delta \ddot{y}(t, n) + \gamma \delta \dot{y}(t, n) + \omega_0^2\delta y(t,
n) + \qquad \qquad \qquad \\
N^{\prime}(y^{odd}(t, n))\delta y(t, n) = h(y^{odd}(t, n))
\end{split}%
\end{equation}

Here $N^{\prime }(y)=\frac{\mathrm{d}N(y)}{\mathrm{d}y}$

One can see that the correction $\delta y(t,n)$ obeys the equation of a
harmonic oscillator driven with a force $h(y^{odd}(t,n))$ and parametrically
driven with $N(y^{odd}(t,n))$. The force $h(y^{odd}(t,n))$ is defined by

\begin{equation}
h(y^{odd}(t,n))=\sum_{k=n+1}^{\infty }\tilde{N}_{2k-1}^{n}(\tilde{y}_{1},...,%
\tilde{y}_{2n-1})\mathrm{e}^{i(2k-1)\omega \tau }+c.c.
\label{appendix_general_correction_force}
\end{equation}

where $c.c.$ is the complex conjugate. Note that $h(y^{odd}(t,n))$ contains
only odd harmonics starting from $2n+1$. If the oscillator is parametrically
stable, the correction $\delta y(t,n)$ will only contain oscillations at
higher odd harmonics. They can be taken into account in principle by
increasing the number of odd harmonics $n$ in Eq.~(\ref{appendix_x_tau_n}).
This means that we can ignore the drive term for the analysis of parametric
instabilities. The stability analysis will now be reduced to the analysis of
the parametrically driven harmonic oscillator which is very well understood.

Because the function $N^{\prime}(y)$ is even $N^{\prime}(y^{odd}(\tau, n))$
will contain only \textit{even} harmonics of $\omega$. That is

\begin{equation}
N^{\prime }(y^{odd}(\tau ,n))=\frac{1}{2}\sum_{k=0}^{\infty }\tilde{%
N^{\prime }}_{2k}^{n}(\tilde{y}_{1},...,\tilde{y}_{2n-1})\mathrm{e}%
^{i2k\omega t}+c.c.  \label{appendix_expansion_of_N'}
\end{equation}

with Fourier amplitudes $\tilde{N^{\prime}}_{2k}^n$ given by

\begin{equation}
\begin{split}
\tilde{N^{\prime }}_{2m}^{n}(z_{1},...,z_{2n-1}) = \\
\int_{-\pi }^{\pi }N^{\prime }\left( \frac{1}{2}\sum_{k=1}^{n}z_{2k-1}%
\mathrm{e}^{i(2k-1)\theta }+c.c.\right) \mathrm{e}^{-i2m\theta }\,\frac{%
d\theta }{\pi }
\end{split}
\label{appendix_tilde_N'}
\end{equation}

Now the equation for the parametric driven correction $\delta y$ can be
rewritten in the form

\begin{equation}
\begin{split}
\delta \ddot{y}(t,n)+\gamma \delta \dot{y}(t,n)+\delta y(t,n)\times \\
\left( 1+\tilde{N^{\prime }}_{0}^{n}+\sum_{k=1}^{\infty }\left( \tilde{%
N^{\prime }}_{2k}^{n}(\tilde{y}_{1},...,\tilde{y}_{2n-1})\mathrm{e}%
^{i2k\omega t}+c.c.\right) \right) =0
\end{split}
\label{appendix_Hills_with_damping}
\end{equation}

Eq.~(\ref{appendix_Hills_with_damping}) is known in the literature on
differential equations as Hill's equation with linear damping. Methods to
investigate its stability diagram both analytically and numerically can be
found in~\cite{Nayfeh} for example.

The goal of our discussion can be achieved by considering the simplest form
of~(\ref{appendix_Hills_with_damping}). We will take $n=1$ corresponding to
the single mode solution $y^{odd}(t,1)\equiv y_{1}(t)=\frac{1}{2}(\tilde{y}%
_{1}e^{i\omega t}+c.c.)$ and truncate the series~(\ref%
{appendix_expansion_of_N'}) to only the first and second terms. This
corresponds to a DC shift in the linear oscillation frequency and to the $%
2\omega $ parametric drive with amplitude $\tilde{N^{\prime }}_{2}^{1}(%
\tilde{y_{1}})$. This approximation is rich enough to understand the
bistability of the first harmonic response of a driven nonlinear oscillator
as discussed in the main body of the paper. It also shows the roads that
lead to the breakdown of the simple picture of bistability. These
simplifications lead to

\begin{equation}
\begin{split}
\delta \ddot{y}+\gamma \delta \dot{y}+\delta y[\omega _{0}^{2}+\tilde{%
N^{\prime }}_{0}^{1}(\tilde{y}_{1})+ \\
|\tilde{N^{\prime }}_{2}^{1}(\tilde{y}_{1})|\cos (2\omega t+Arg[\tilde{%
N^{\prime }}_{2}^{1}])] =0
\end{split}
\label{appendix_Mathieu_with_damping}
\end{equation}

where we have simplified our notation by using $\delta y(t,1)\equiv \delta y$%
. To reach the canonical form let us shift and rescale time by defining $%
2t^{\prime }\equiv 2\omega t+Arg[\tilde{N^{\prime }}_{2}^{1}]$ and introduce
two parameters of central importance:

\begin{equation}
\alpha ={\frac{\omega _{0}^{2}+\tilde{N^{\prime }}_{0}^{1}}{\omega ^{2}}}%
;\qquad \beta ={\frac{|\tilde{N^{\prime }}_{2}^{1}|}{\omega ^{2}}}
\label{appendix_alpha_beta}
\end{equation}

where $\tilde{N^{\prime }}$ is evaluated at $\tilde{y}_{1}$. We get%
\begin{equation}
\delta \ddot{y}+\frac{1}{Q}\delta y+(\alpha +\beta \cos 2t^{\prime })\delta
y=0  \label{appendix_Mathieu}
\end{equation}

this is known as Mathieu's equation with damping~\cite{Nayfeh}. The main
instability region corresponds to \textit{order one parametric resonance}.
It can be intuitively understood as a result of efficient pumping of an
oscillator at the frequency $\sqrt{\alpha }\sim 1$. This instability leads
to growing oscillations of $\delta y\sim \mathrm{e}^{\lambda ^{\prime
}t^{\prime }}\mathrm{e}^{it^{\prime }}=\mathrm{e}^{\lambda \tau }\mathrm{e}%
^{i\omega t}$. Importantly, there are at \textit{the same frequency as our
odd anzatz }$y^{odd}(t,1)$. This can be incorporated into the anzatz by
replacing $\tilde{y}_{1}\rightarrow \tilde{y}_{1}(t)$. This was done in the
main body of the paper and was shown to lead to hysteresis and bistability.
In terms of $\alpha (\omega ,\tilde{y}_{1})$ and $\beta (\omega ,\tilde{y}%
_{1})$ the unstable (bistable) region is given by

\begin{equation}  \label{appendix_parametric_resonance1}
\alpha(\omega, \tilde{y}_1) = 1 \pm \sqrt{\frac{\beta(\omega, \tilde{y}_1)^2%
}{4} - \frac{1}{Q^2}} - \frac{\beta(\omega, \tilde{y}_1)^2}{32}
\end{equation}

The two nearest instabilities of the "wrong" type correspond to a second
order parametric resonance when $\sqrt{\alpha }\approx 2$ and to the
phenomenon of the drive-mediated negative restoring force when $\alpha <0$.
These two instabilities contain growing double-frequency and zero-frequency
components in $\delta y(t)$, respectively and they break the symmetry of the
non-driven problem. The locus of these transitions, with an accuracy of
order $o(|\beta (\omega ,\tilde{y}_{1})|^{2})$ is given respectively by

\begin{equation}  \label{appendix_parametric_resonance2}
\alpha(\omega, \tilde{y}_1) = 4 + \frac{1}{24} \beta(\omega, \tilde{y}_1)^2
\pm \sqrt{\frac{\beta(\omega, \tilde{y}_1)^2}{128} - \frac{4}{Q^2}}
\end{equation}
and
\begin{equation}  \label{appendix_negative_spring}
\alpha(\omega, \tilde{y}_1) = - \frac18 \beta(\omega, \tilde{y}_1)^2
\end{equation}

For the Duffing potential, where $N(y)=\omega _{0}^{2}y^{3}$ we found that $%
\alpha $ and $\beta $ have particularly simple form: $\alpha ^{D}={\frac{%
\omega _{0}^{2}}{\omega ^{2}}(1-\frac{3}{2}}\tilde{y}{_{1}^{2})}$ and $\beta
^{D}={-\frac{\omega _{0}^{2}}{\omega ^{2}}\frac{3}{2}}\tilde{y}{_{1}^{2}}$.

We now arrive at the main result of this Appendix: provided that $|\beta
^{D}|={\frac{\omega _{0}^{2}}{\omega ^{2}}\frac{3}{2}}\tilde{y}{_{1}^{2}<1}$
and $1/2<w=\omega /\omega _{0}<1$ the symmetry breaking unstable regimes are
inaccessible since neither of Eqs.~(\ref{appendix_parametric_resonance2}),~(%
\ref{appendix_negative_spring}) has a solution. The analysis conducted in
this Appendix provides the ground for discussing the nonlinear oscillator
only in terms of a single frequency oscillating response. We saw in the main
body of the paper how this leads to a simple picture of bistability. The
exact boundary at which this picture breaks down requires a sophisticated
analysis. However the simple analysis reveals that the instability regions
are well separated from our region of interest for the relevant experimental
parameters ($1-w\sim 1/Q\ll 1$, $|\tilde{y}_{1}|\sim 1/\sqrt{Q}\ll 1$). More
importantly, it shows that the breakdown of the simple bifurcation picture
is $Q$-independent in the limit of high $Q$ and the size of the accessible
bistability region can therefore be under control.

In conclusion, basing ourselves on references~\cite{Bogolyubov_Mitropolskii,
Duffing_stability1, Duffing_stability2, Duffing_stability3, Nayfeh} we have
explained that instabilities of the steady state response of a Josephson
junction nonlinear oscillator can be viewed as parametric resonances. This
mapping allows to classify the different instabilities and separate the ones
of interest for bifurcation amplification from the unwanted ones. A "sound"
bifurcation is garanteed when i) the quality factor is much greater than
unity, ii) the relative detuning is of the order of the inverse of the
quality factor and iii) the dimensionless oscillation amplitude is of order
of the inverse of the square of the quality factor.



\bigskip

\begin{thebibliography}{99}
\bibitem{HEMT} N. Wadefalk, A. Mellberg, I. Angelov, M. E. Barsky, S. Bui,
E. Choumas, R. W. Grundbacher, E. L. Kollberg, R. Lai, N. Rorsman, P.
Starski, J. Stenarson, D. C. Streit, H. Zirath \textit{et. al.}, IEEE Trans.
on Microwave Theory and Tech. \textbf{51}, 1705 (2003); N. Oukhanski, M.
Grajcar. E. Il'ichev, H.-G. Meyer, Rev. Sci. Instrum. \textbf{74}, 1145
(2003).

\bibitem{Squid_amp_darkmatter} R. Bradley, J. Clarke, D. Kinion, L.
Rosenberg , K. van Bibber, S. Matsuki, M. M\"{u}ck, P. Sikivie, Rev. Mod.
Phys. \textbf{75}, 777 (2003).

\bibitem{JBA} I. Siddiqi, R. Vijay, F. Pierre, C. M. Wilson, M. Metcalfe, C.
Rigetti, L. Frunzio, M. H. Devoret, Phys. Rev. Lett. \textbf{93}, 207002
(2004).

\bibitem{Dynamical_switching_of_JJ} I. Siddiqi, R. Vijay, F. Pierre, C. M.
Wilson, L. Frunzio, M. Metcalfe, C. Rigetti, R. J. Schoelkopf, M. H.
Devoret, D. Vion, D. Esteve Phys. Rev. Lett. \textbf{94}, 027005 (2005).

\bibitem{JBA_and_qubit} I. Siddiqi, R. Vijay, M. Metcalfe, E. Boaknin, L.
Frunzio, R. J. Schoelkopf, M. H. Devoret, Phys. Rev. B \textbf{73}, 054510
(2006).

\bibitem{SQUID} M. M\"{u}ck, C. Welzel, J. Clarke, Appl. Phys. Lett. \textbf{%
82}, 3266 (2003).

\bibitem{SSET} R. J. Schoelkopf, P. Wahlgren, A. A. Kozhevnikov, P. Delsing,
D. E. Prober, Science \textbf{280}, 1238 (1998).

\bibitem{Yurke_review} B. Yurke, L. R. Corruccini, P. G. Kaminsky, L. W.
Rupp, A. D. Smith, A. H. Silver, R. W. Simon, E. A Whittaker, Phys. Rev. A%
\textbf{\ 39}, 2519 (1989).

\bibitem{RF_SQUID} L. D. Jackel and R. A. Buhrman, J. Low Temp. Phys.
\textbf{19}, 210 (1974).

\bibitem{Bogolyubov_Mitropolskii} N. N. Boglyubov and Y. A. Mitropolskii
"Asymptotic methods is the theory of non-linear oscillations" (Gordon and
Breach, New York, 1961).

\bibitem{Nayfeh} A. H. Nayfeh and D. T. Mook, "Nonlinear oscillations"
(John Wiley~$\&$~Sons, New York 1979).

\bibitem{Landau_Lifshitz} L. D. Landau and E. M Lifshitz, "Mechanics"
(Pergamon, Oxford, 1969).

\bibitem{Dykman_Krivoglaz80} M. I. Dykman and M. A. Krivoglaz, Physica A
\textbf{104}, 480-494, (1980).

\bibitem{susceptibility_footnote} In some instances, it may be more relevant
to look at the maximum susceptibility $\partial {\bf x}/\partial
\Omega $ with respect to $\Omega $. For a positive susceptibility
(on the low frequency
side), we can define a similar line $f_{ms}^{\Omega }$. It is given by $%
\frac{f_{ms}(\Omega )}{f_{c} }=\frac{1}{\sqrt{2}}\left( 3\frac{\Omega }{%
\Omega _{c}}-1\right) ^{1/2}$

\bibitem{Duffing_stability1} Y. H. Kao, J. C. Huang, and Y. S. Gou, Phys.
Lett. A, \textbf{131}, 92 (1988).

\bibitem{Duffing_stability2} S. Novak, R. G. Frehlich, Phys. Rev. A \textbf{%
26}, 3660-3663 (1982).

\bibitem{Duffing_stability3} B. A. Huberman and J. P. Crutchfield, Phys.
Rev. Lett. \textbf{43}, 1743 (1979).

\bibitem{Josephson_relations} B. D. Josephson, Rev. Mod. Phys, \textbf{36},
216, (1964).

\bibitem{Zmuid} P. K. Day, H. G. LeDuc, B. A. Mazin, A. Vayonakis, J.
Zmuidzinas, Nature \textbf{425}, 817 (2003).

\bibitem{Wallraff} A. Wallraff, D. I. Schuster, A. Blais, L. Frunzio, R.-S.
Huang, J. Majer, S. Kumar, S. M. Girvin and R. J. Schoelkopf, Nature \textbf{%
431}, 162 (2004)

\bibitem{Pozar} D.M. Pozar, \textit{Microwave Engineering}, (John Wiley and
Sons, Hoboken, 2005).

\bibitem{janicelee} J. C. Lee, W. D. Oliver, T. P. Orlando, K. K. Berggren,
IEEE Trans. Appl. Supercon. \textbf{15}, 841 (2005).

\bibitem{lupascu} A. Lupascu, E. F. C. Driessen, L. Roschier, C. J. P. M.
Harmans, J. E. Mooij, Phys. Rev. Lett. \textbf{96}, 127003 (2006).

\bibitem{Dykman_Krivoglaz79} M. I. Dykman and M. A. Krivoglaz, Zh. Eksp.
Teor. Fiz. \textbf{77}, 60-73, (1979).

\bibitem{Dykman_Smelyanski88} M. I. Dykman and V. N. Smelyanskii, Zh. Eksp.
Teor. Fiz., \textbf{94}, 61-74, (1988).

\bibitem{Dykman} M. I. Dykman, cond-mat 0606198

\end{thebibliography}
\end{document}